\documentclass[prl,showpacs,twocolumn]{revtex4}
\usepackage{times}
\usepackage{amsmath}
\usepackage{graphicx}
\usepackage{epsfig}
\usepackage{dcolumn}
\usepackage{bm}
\usepackage{color}
\newcommand{\BR}{{\cal B}}
\newcommand{\pp}{\pi^+\pi^-}
\newcommand{\pip}{\pi^+}
\newcommand{\pim}{\pi^-}
\newcommand{\piz}{\pi^0}

\newcommand{\LL}{\ell^+\ell^-}
\newcommand{\EE}{e^+e^-}
\newcommand{\MM}{\mu^+\mu^-}

\newcommand{\psip}{\psi(3686)}

\newcommand{\jpsi}{J/\psi}

\newcommand{\pcpcjpsi}{\pi^+\pi^-J/\psi}

\newcommand{\y}{Y(4260)}
\newcommand{\X}{Z_c(3900)}
\newcommand{\psithr}{\psi(4040)}
\newcommand{\psifou}{\psi(4160)}
\newcommand{\psifiv}{\psi(4415)}
\newcommand{\ddb}{D\bar{D}}

\newcommand{\bbb}{B\bar{B}}

\def\Journal#1#2#3#4{{#1} {\bf #2}, #3 (#4)}

\def\PLB{Phys. Lett. B}
\def\PRL{Phys. Rev. Lett.}
\def\PRD{Phys. Rev. D}

\def\EPJC{Eur. Phys. J. C}

\begin{document}


\title{\boldmath
Observation of a charged charmoniumlike structure in $\EE\to
\pp\jpsi$ at $\sqrt{s}=4.26$~GeV}

\author{{
M.~Ablikim$^{1}$, M.~N.~Achasov$^{6}$, X.~C.~Ai$^{1}$,
O.~Albayrak$^{3}$, D.~J.~Ambrose$^{39}$, F.~F.~An$^{1}$,
Q.~An$^{40}$, J.~Z.~Bai$^{1}$, R.~Baldini Ferroli$^{17A}$,
Y.~Ban$^{26}$
% ...
\\ \vspace{0.2cm}
(BESIII Collaboration)\\ \vspace{0.2cm}
 {\it
$^{1}$ Institute of High Energy Physics, Beijing 100049, People's Republic of China\\
% ...
$^{3}$ Carnegie Mellon University, Pittsburgh, Pennsylvania 15213, USA\\
% ...
$^{6}$ G.I. Budker Institute of Nuclear Physics SB RAS (BINP), Novosibirsk 630090, Russia\\
% ...
$^{17}$ (A)INFN Laboratori Nazionali di Frascati, I-00044, Frascati, Italy; (B)INFN and University of Perugia, I-06100, Perugia, Italy\\
% ...
$^{26}$ Peking University, Beijing 100871, People's Republic of China\\
% ...
$^{39}$ University of Rochester, Rochester, New York 14627, USA\\
$^{40}$ University of Science and Technology of China, Hefei 230026, People's Republic of China\\
% ...
}}}

\date{\today}

\begin{abstract}

We study the process $\EE\to \pcpcjpsi$ at a center-of-mass energy
of $4.260$~GeV using a 525~pb$^{-1}$ data sample collected with
the BESIII detector operating at the Beijing Electron Positron
Collider. The Born cross section is measured to be $(62.9\pm
1.9\pm 3.7)$~pb, consistent with the production of the $\y$. We
observe a structure at around 3.9~GeV/$c^2$ in the $\pi^\pm \jpsi$
mass spectrum, which we refer to as the $\X$. If interpreted as a
new particle, it is unusual in that it carries an electric charge
and couples to charmonium. A fit to the $\pi^\pm\jpsi$ invariant
mass spectrum, neglecting interference, results in a mass of
$(3899.0\pm 3.6\pm 4.9)~{\rm MeV}/c^2$ and a width of $(46\pm
10\pm 20)$~MeV. Its production ratio is measured to be
$R=\frac{\sigma(\EE\to \pi^\pm \X^\mp\to \pcpcjpsi))}
{\sigma(\EE\to \pcpcjpsi)}=(21.5\pm 3.3\pm 7.5)\%$. In all
measurements the first errors are statistical and the second are
systematic.

\end{abstract}


\maketitle


Since its discovery in the initial-state-radiation (ISR) process
$\EE \to \gamma_{\rm ISR}\pcpcjpsi$~\cite{babay4260}, and despite
its subsequent
observations~\cite{cleo_y,belley4260,cleoy4260,babarnew}, the
nature of the $\y$ state has remained a mystery. Unlike other
charmonium states with the same quantum numbers and in the same
mass region, such as the $\psithr$, $\psifou$, and $\psifiv$, the
$\y$ state does not have a natural place within the quark model of
charmonium~\cite{qm}.  Furthermore, while being well above the
$\ddb$ threshold, the $\y$ shows strong coupling to the
$\pcpcjpsi$ final state~\cite{moxh_y}, but relatively small
coupling to open charm decay
modes~\cite{dd16,dd22,dd27,dd28,dd29}. These properties perhaps
indicate that the $\y$ state is not a conventional state of
charmonium~\cite{epjc-review}.

A similar situation has recently become apparent in the
bottomonium system above the $\bbb$ threshold, where there are
indications of anomalously large couplings between the
$\Upsilon(5S)$ state (or perhaps an unconventional bottomonium
state with similar mass, the $Y_b(10890)$) and the
$\pp\Upsilon(1S,2S,3S)$ and $\pp h_b(1P,2P)$ final
states~\cite{yb1,yb2}.  More surprisingly, substructure in these
$\pp\Upsilon(1S,2S,3S)$ and $\pp h_b(1P,2P)$ decays indicates the
possible existence of charged bottomoniumlike states~\cite{zb},
which must have at least four constituent quarks to have a
non-zero electric charge, rather than the two in a conventional
meson. By analogy, this suggests there may exist interesting
substructure in the $\y\to \pcpcjpsi$ process in the charmonium
region.

In this Letter, we present a study of the process $\EE\to
\pcpcjpsi$ at a center-of-mass (CM) energy of  $\sqrt{s}=
(4.260\pm 0.001)$~GeV, which corresponds to the peak of the $\y$
cross section. We observe a charged structure in the
$\pi^\pm\jpsi$ invariant mass spectrum, which we refer to as the
$\X$. The analysis is performed with a 525~pb$^{-1}$ data sample
collected with the BESIII detector, which is described in detail in Ref.~\cite{bepc2}.
In the studies presented here, we rely only on charged particle
tracking in the main drift chamber~(MDC) and energy deposition
in the electromagnetic calorimeter~(EMC).

The {\sc geant4}-based Monte Carlo~(MC) simulation software, which
includes the geometric description of the BESIII detector and the
detector response, is used to optimize the event selection
criteria, determine the detection efficiency, and estimate
backgrounds. For the signal process, we use a sample of $\EE\to
\pcpcjpsi$ MC events generated assuming the $\pcpcjpsi$ is
produced via $\y$ decays, and using the $\EE\to \pcpcjpsi$ cross
sections measured by Belle~\cite{belley4260} and
BaBar~\cite{babarnew}. The $\pcpcjpsi$ substructure is modelled
according to the experimentally observed Dalitz plot distribution
presented in this analysis. ISR is simulated with {\sc
kkmc}~\cite{kkmc} with a maximum energy of 435~MeV for the ISR
photon, corresponding to a $\pcpcjpsi$ mass of 3.8~GeV/$c^2$.


For $\EE\to \pcpcjpsi$ events, the $\jpsi$ candidate is
reconstructed with lepton pairs ($\EE$ or $\MM$).  Since this
decay results in a final state with four charged particles, we
first select events with four good charged tracks with net charge
zero. For each charged track, the polar angle in the MDC must
satisfy $|\cos\theta|<0.93$, and the point of closest approach to
the $\EE$ interaction point must be within $\pm 10$~cm in the beam
direction and within $1$~cm in the plane perpendicular to the beam
direction. Since pions and leptons are kinematically well
separated in this decay, charged tracks with momenta larger than
1.0~GeV/$c$ in the lab frame are assumed to be leptons, and the
others are assumed to be pions. We use the energy deposited in the
EMC to separate electrons from muons. For muon candidates, the
deposited energy in the EMC should be less than 0.35~GeV; while
for electrons, it should be larger than 1.1~GeV. The efficiencies
of these requirements are determined from MC simulation to be
above 99\% in the EMC sensitive region.

In order to reject radiative Bhabha and radiative dimuon
($\gamma\EE/\gamma\MM$) backgrounds associated with a
photon-conversion, the cosine of the opening angle of the pion
candidates, which are true $\EE$ pairs in the case of background,
is required to be less than 0.98. In the $\EE$ mode, the same
requirement is imposed on the $\pi^{\pm}e^{\mp}$ opening angles.
This restriction removes less than 1\% of the signal events.

The lepton pair and the two pions are subjected to a
four-constraint~(4C) kinematic fit to the total initial
four-momentum of the colliding beams in order to improve the
momentum resolution and reduce the background. The $\chi^2$ of the
kinematic fit is required to be less than 60.

\begin{figure*}
\begin{center}
\includegraphics[width=0.45\textwidth]{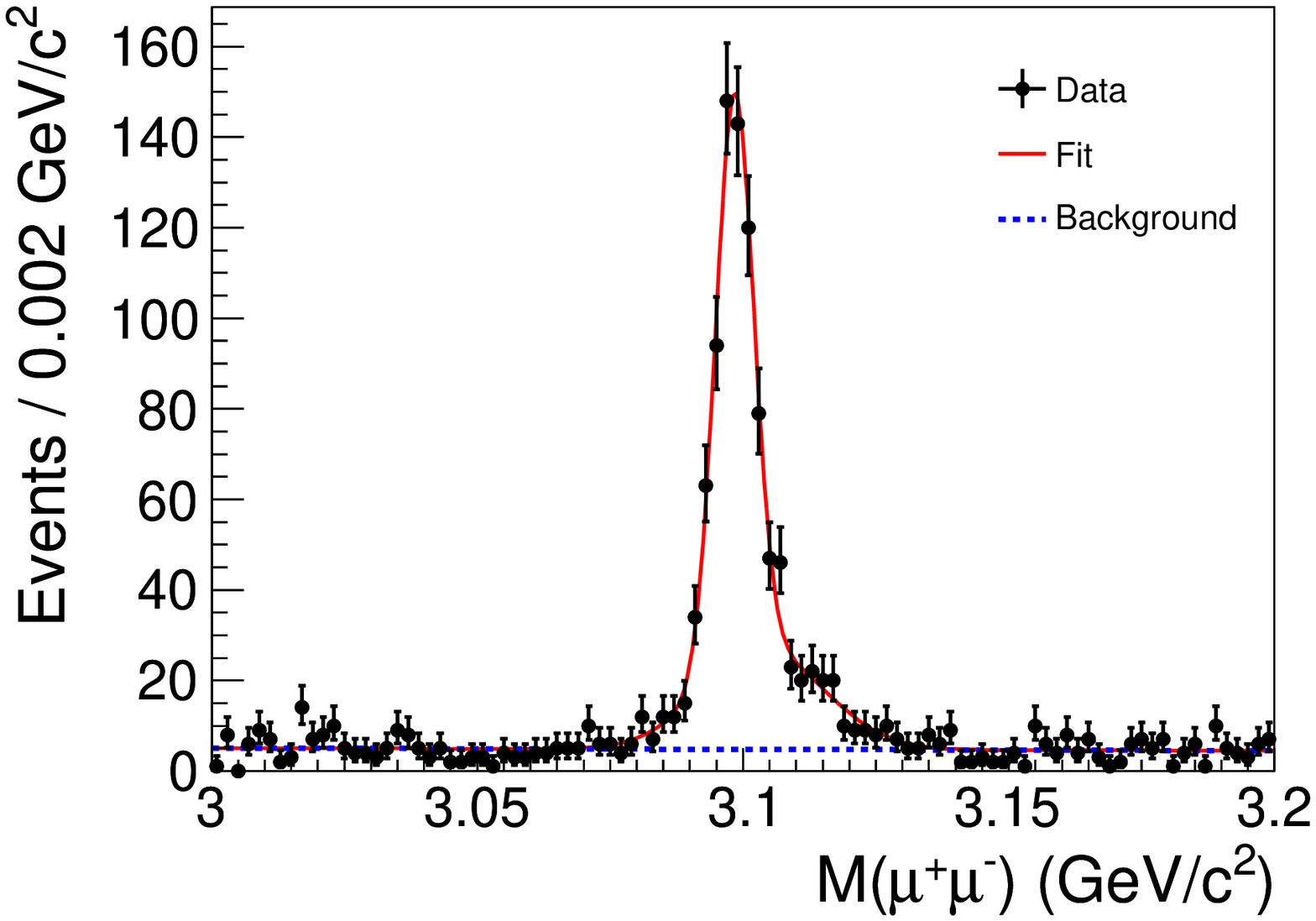}
\includegraphics[width=0.45\textwidth]{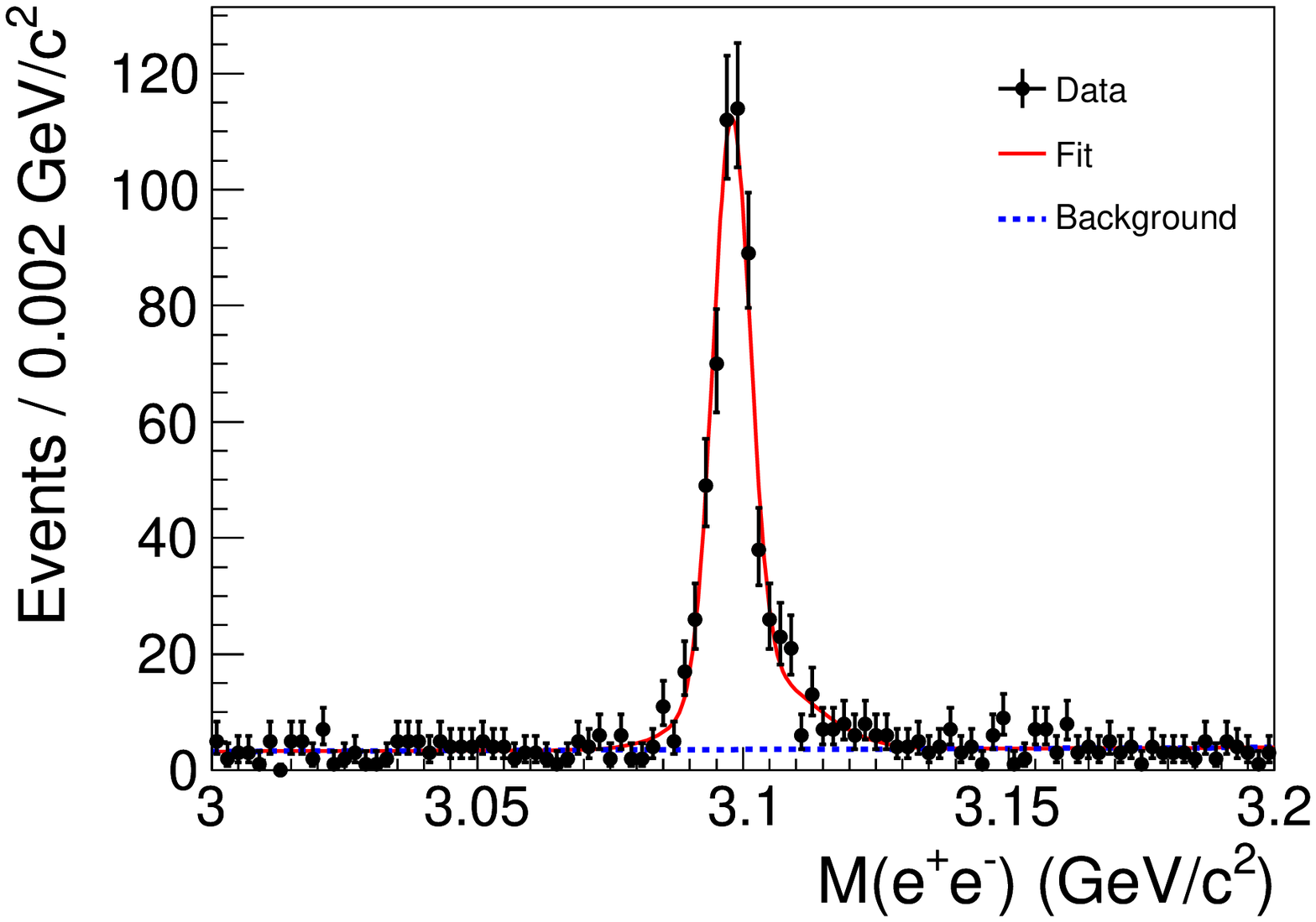}
\caption{The distributions of $M(\MM)$ (left panel) and $M(\EE)$
(right panel) after performing a 4C kinematic fit and imposing all
selection criteria. Dots with error bars are data and the curves
are the best fit described in the text.} \label{m2l}
\end{center}
\end{figure*}

After imposing these selection criteria, the invariant mass
distributions of the lepton pairs are shown in Fig.~\ref{m2l}. A
clear $\jpsi$ signal is observed in both the $\EE$ and $\MM$
modes. There are still remaining $\EE\to \pp\pp$, and other QED
backgrounds, but these can be estimated using the events in the
$\jpsi$ mass sideband.  The final selection efficiency is $53.8\pm
0.3$\% for $\MM$ events and $38.4\pm 0.3$\% for $\EE$ events,
where the errors are from the statistics of the MC sample.
The main factors affecting the detection efficiencies include the detector acceptances for four charged tracks and
the requirement on the quality of the kinematic fit adopted. The lower efficiency for $\EE$ events is due to
final-state-radiation (FSR), bremsstrahlung energy loss of $\EE$ pairs and the EMC deposit energy requirement.

To extract the number of $\pcpcjpsi$ signal events, invariant mass
distributions of the lepton pairs are fit using the sum of two
Gaussian functions with a linear background term. The fits yield
$M(\jpsi)= 3098.4\pm 0.2$~MeV/$c^2$ with $882\pm 33$ signal events
in the $\MM$ mode; and $M(\jpsi)=3097.9\pm 0.3$~MeV/$c^2$ with
$595\pm 28$ signal events in the $\EE$ mode. Here the errors are
statistical only. The mass resolution is 3.7~MeV/$c^2$ in the
$\MM$ mode and 4.0~MeV/$c^2$ in the $\EE$ mode.


The Born cross section is determined from the relation
 $
   \sigma^{B}=\frac{N^{\rm fit}}
   {\mathcal{L}_{\rm int}(1+\delta) \epsilon \BR},
 $
where $N^{\rm fit}$ is the number of signal events from the fit;
$\mathcal{L}_{\rm int}$ is the integrated luminosity; $\epsilon$
is the selection efficiency obtained from a MC simulation; $\BR$
is the branching fraction of $\jpsi\to \LL$; and ($1+\delta$) is
the radiative correction factor, which is $0.818$ according to a
QED calculation~\cite{kuraev}. The measured Born cross section for
$\EE\to \pcpcjpsi$ is $(64.4\pm 2.4)$~pb in the $\MM$ mode and
$(60.7\pm 2.9)$~pb in the $\EE$ mode. The combined measurement is
$\sigma^B(\EE\to\pcpcjpsi)=(62.9\pm 1.9)$~pb.

Systematic errors in the cross section measurement come from the
luminosity measurement, tracking efficiency, kinematic fit,
background estimation, dilepton branching fractions of the
$\jpsi$, and $\y$ decay dynamics.

%

The integrated luminosity of this data sample was measured using
large angle Bhabha events, and has an estimated uncertainty of
1.0\%. The tracking efficiency uncertainty is estimated to be 1\% for each
track from a study of the control samples $\jpsi\to \pp\piz$ and
$\psip\to \pcpcjpsi$. Since the luminosity is measured using
Bhabha events, the tracking efficiency uncertainty of high momentum lepton
pairs partly cancels in the calculation of the $\pcpcjpsi$ cross
section. To be conservative, we take 4\% for both the $\EE$ and
$\MM$ modes.

The uncertainty from the kinematic fit comes from the
inconsistency between the data and MC simulation of the track
helix parameters. Following the procedure described in
Ref.~\cite{guoyp}, we take the difference between the efficiencies
with and without the helix parameter correction as the systematic
error, which is 2.2\% in the $\MM$ mode and 2.3\% in the $\EE$
mode.

Uncertainties due to the choice of background shape and fit range
are estimated by varying the background function from linear to a
second-order polynomial and by extending the fit range.

Uncertainties in the $\y$ resonance parameters and possible
distortions of the $\y$ line shape introduce small systematic
uncertainties in the radiative correction factor and the
efficiency. This is estimated using the different line shapes
measured by Belle~\cite{belley4260} and BaBar~\cite{babarnew}. The
difference in $(1+\delta)\cdot\epsilon$ is 0.6\% in both the $\EE$
and $\MM$ modes, and this is taken as a systematic error.

We use the observed Dalitz plot to generate $\y\to \pcpcjpsi$
events. To cover possible modelling inaccuracies, we
conservatively take the difference between the efficiency using
this model and the efficiency using a phase space model as a
systematic error.  The error is 3.1\% in both the $\MM$ and the
$\EE$ modes.

The uncertainty in $\BR(\jpsi\to \LL)$ is 1\%~\cite{PDG}. The
trigger simulation, the event start time determination, and the
final-state-radiation simulation are well-understood; the total
systematic error due to these sources is estimated to be less than
1\%.

Assuming all of the sources are independent, the total systematic
error in the $\pcpcjpsi$ cross section measurement is determined
to be 5.9\% for the $\MM$ mode and 6.8\% for the $\EE$ mode.
Taking the correlations in errors between the two modes into
account, the combined systematic error is slightly less than
5.9\%.


Intermediate states are studied by examining the Dalitz plot of
the selected $\pcpcjpsi$ candidate events. The $\jpsi$ signal is
selected using $3.08 < M(\LL) < 3.12$~GeV/$c^2$ and the sideband
using $3.00 < M(\LL) < 3.06$~GeV/$c^2$ or $3.14 < M(\LL) <
3.20$~GeV/$c^2$, which is three times the size of the signal
region. In total, a sample of 1595 $\pcpcjpsi$ events with a
purity of 90\% is obtained.

\begin{figure}
\begin{center}
\includegraphics[width=0.45\textwidth]{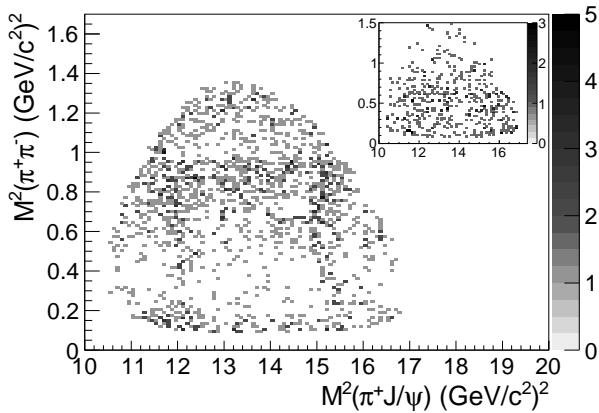}
\caption{Dalitz distributions of $M^2(\pp)$ vs. $M^2(\pi^+\jpsi)$
for selected $\EE\to \pcpcjpsi$ events in the $\jpsi$
signal region. The inset shows background events from the
$\jpsi$ mass sidebands~(not normalized).}
\label{dalitz}
\end{center}
\end{figure}

Figure~\ref{dalitz} shows the Dalitz plot of events in the
$\jpsi$ signal region, where there are structures in the $\pp$ system and
evidence for an exotic charmoniumlike structure in the
$\pi^\pm\jpsi$ system. The inset shows background events from
$\jpsi$ mass sidebands~(not normalized), where no obvious structures are observed.

\begin{figure*}
\begin{center}
\includegraphics[width=0.32\textwidth]{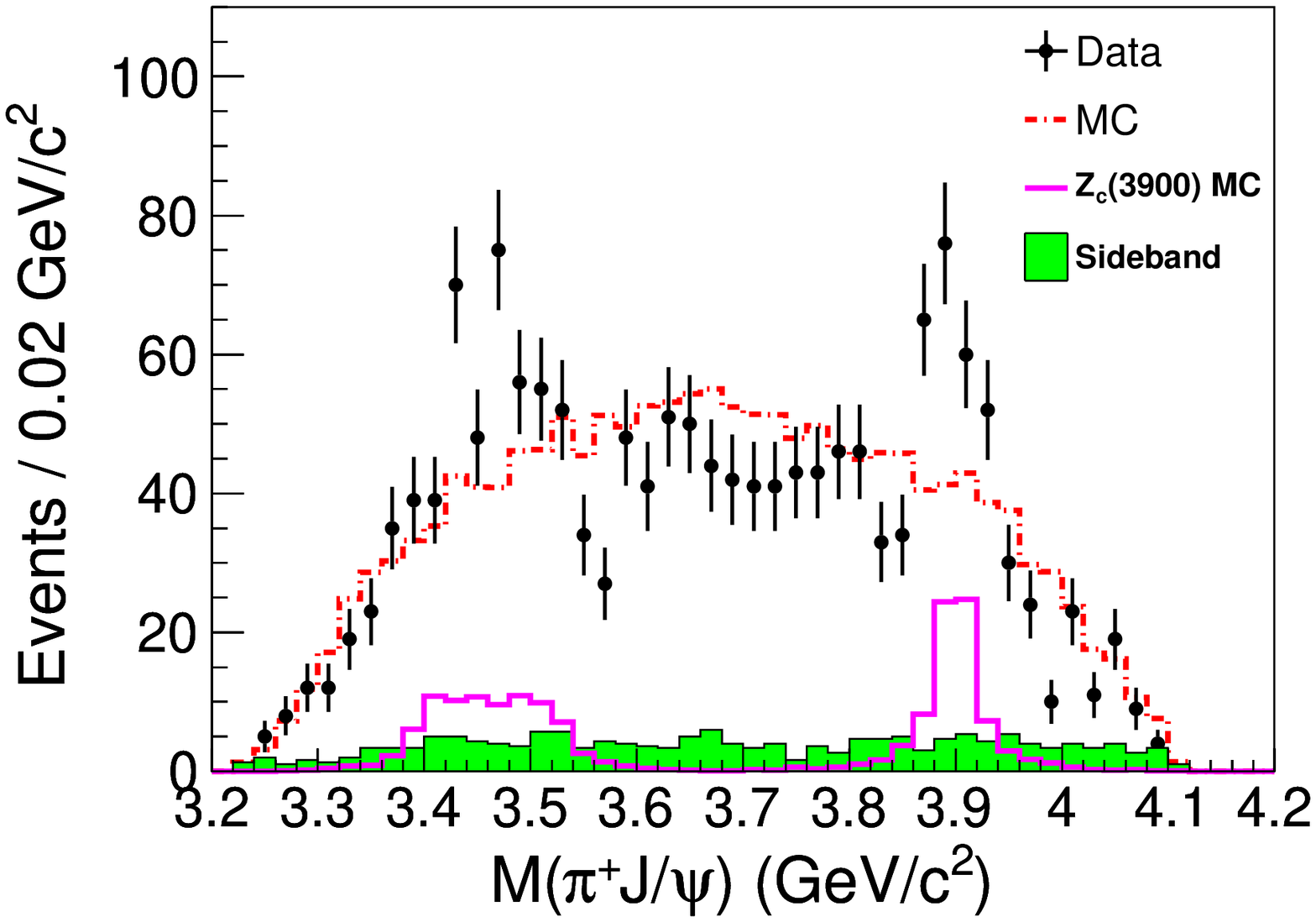}
\includegraphics[width=0.32\textwidth]{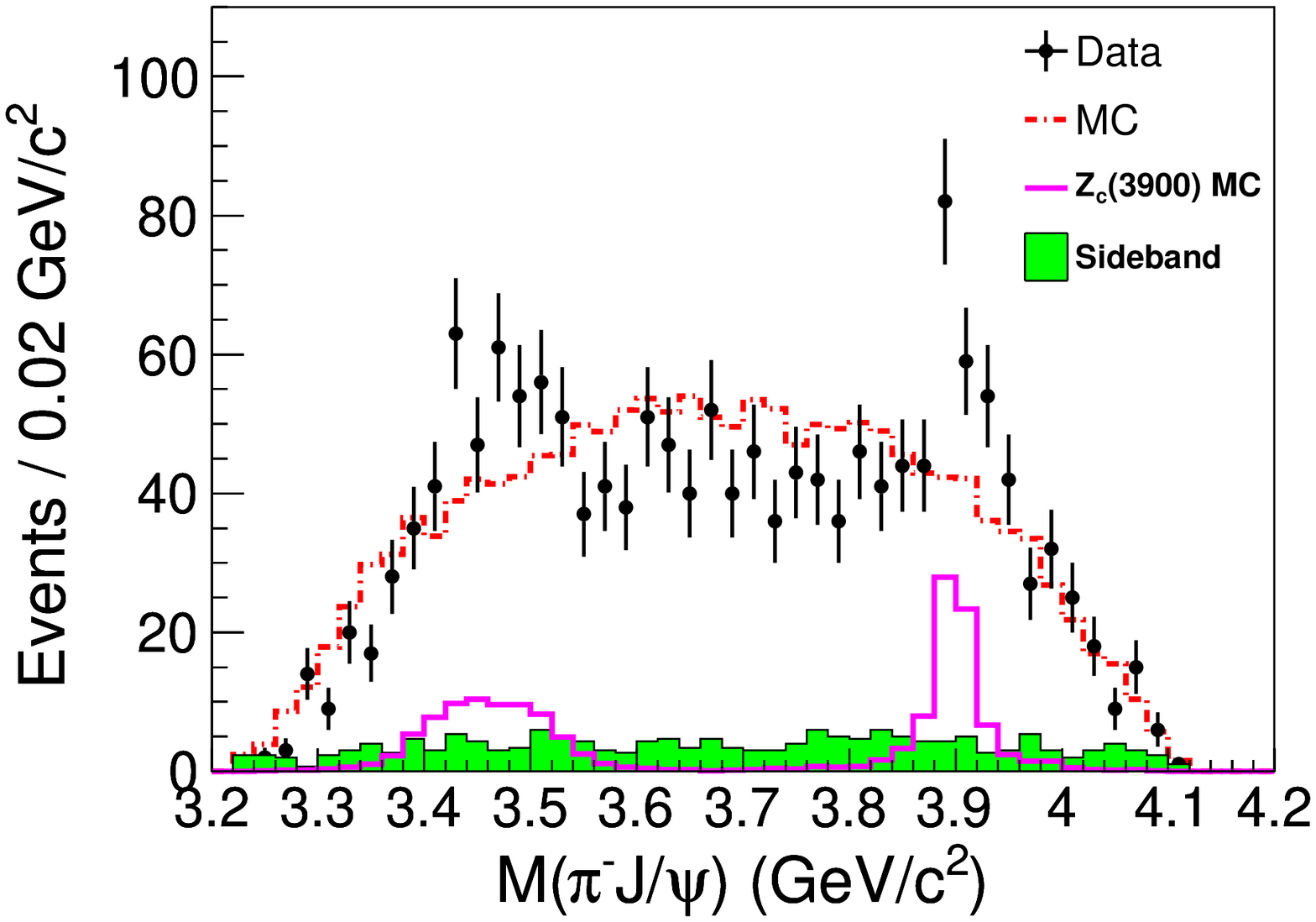}
\includegraphics[width=0.32\textwidth]{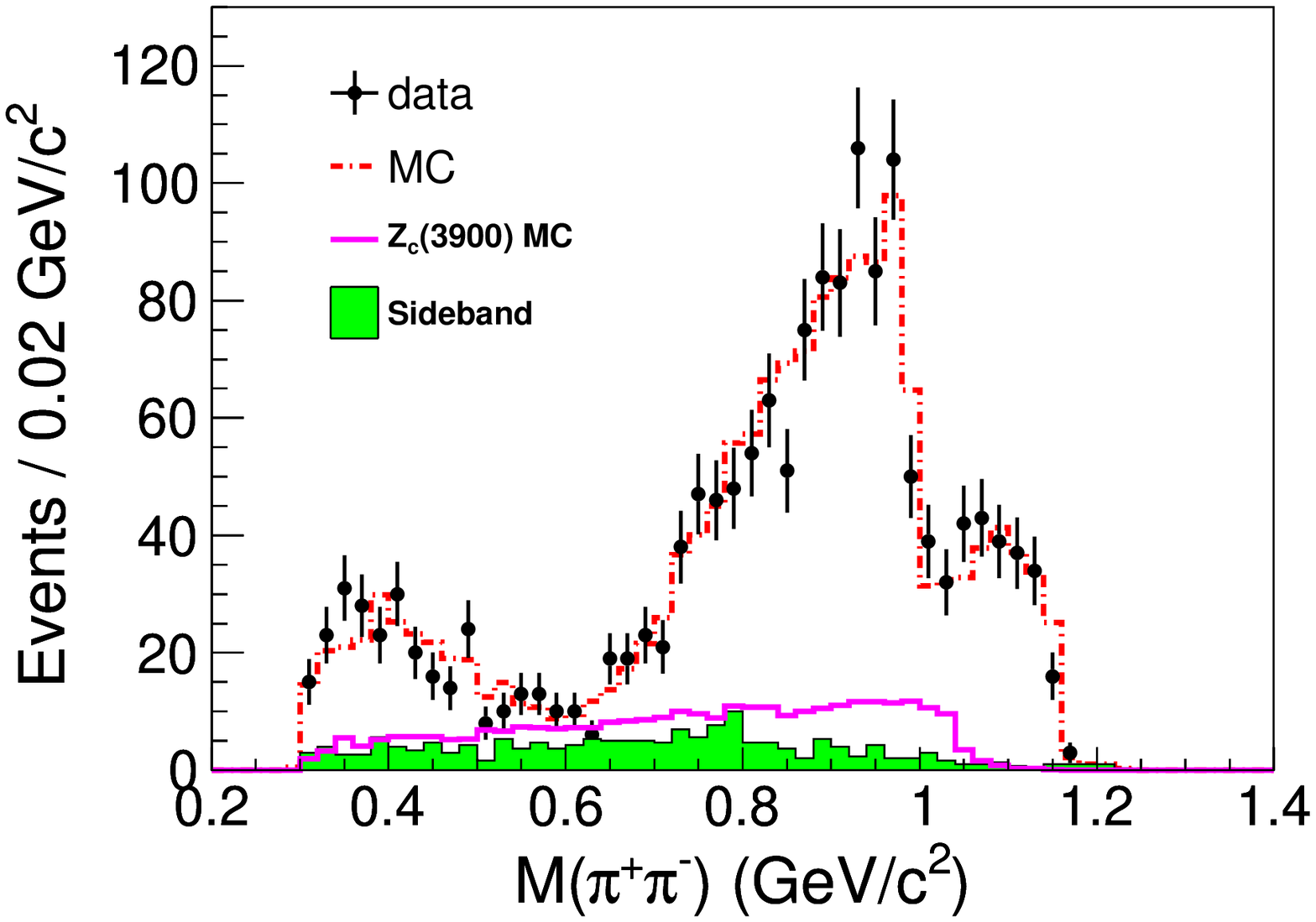}
\caption{One dimensional projections of the $M(\pi^+\jpsi)$,
$M(\pi^-\jpsi)$, and $M(\pp)$ invariant mass distributions in
$\EE\to \pcpcjpsi$ for data in the $\jpsi$ signal region (dots
with error bars), data in the $\jpsi$ sideband region (shaded
histograms), and MC simulation results from $\sigma(500)$, $f_0(980)$
and non-resonant $\pp$ amplitudes (red dot-dashed
histograms). The pink blank histograms show a MC simulation of the
$\X$ signal with arbitrary normalization.} \label{proj}
\end{center}
\end{figure*}

Figure~\ref{proj} shows the projections of the $M(\pi^+\jpsi)$,
$M(\pi^-\jpsi)$, and $M(\pp)$ distributions for the signal events,
as well as the background events estimated from normalized $\jpsi$
mass sidebands.  In the $\pi^\pm\jpsi$ mass spectrum, there is a
significant peak at around 3.9~GeV/$c^2$ (referred to as the $\X$
hereafter). The wider peak at low mass is a reflection of the $\X$
as indicated from MC simulation, and shown in Fig.~\ref{proj}.
Similar structures are observed in the $\EE$ and $\MM$ separated
samples.


The $\pp$ mass spectrum shows non-trivial structure. To
test the possible effects of dynamics in the $\pp$ mass spectrum
on the $\pi^\pm\jpsi$ projection, we develop a parameterization
for the $\pp$ mass spectrum that includes a $f_0(980)$, $\sigma(500)$,
and a non-resonant amplitude. An MC sample generated with
this parameterization adequately describes the $\pp$ spectrum,
as shown in Fig.~\ref{proj}, but does not generate any peaking structure
in the $\pi^\pm\jpsi$ projection consistent with the $\X$.  We have also
tested $D$-wave $\pp$ amplitudes, which are not apparent in the data,
and they, also, do not generate peaks in the $\pi^\pm\jpsi$ spectrum.

\begin{figure}
\begin{center}
\includegraphics[width=0.45\textwidth]{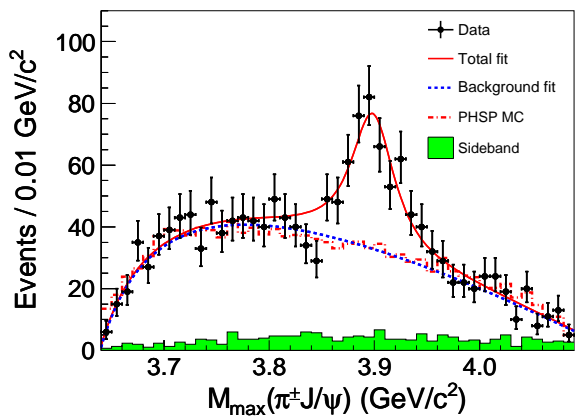}
\caption{Fit to the $M_{\rm max}(\pi^\pm\jpsi)$ distribution as
described in the text. Dots with error bars are data; the red
solid curve shows the total fit, and the blue dotted curve the
background from the fit; the red dot-dashed histogram shows the
result of a phase space MC simulation; and the green shaded
histogram shows the normalized $\jpsi$ sideband events.}
\label{1Dfit}
\end{center}
\end{figure}

An unbinned maximum likelihood fit is applied to the distribution
of $M_{\rm max}(\pi^\pm\jpsi)$, the larger one of the
two mass combinations $M(\pip\jpsi)$ and $M(\pim\jpsi)$ in each event.
The signal shape is parameterized as an S-wave
Breit-Wigner~(BW) function convolved with a Gaussian with a mass
resolution fixed at the MC simulated value (4.2~MeV/$c^2$). The
phase space factor $p\cdot q$ is considered in the partial width, where
$p$ is the $\X$ momentum in the $\y$ CM frame and $q$ is the
$\jpsi$ momentum in the $\X$ CM frame. The background shape is
parameterized as $a/(x-3.6)^b+c+dx$, where $a$, $b$, $c$, and $d$
are free parameters and $x=M_{\rm max}(\pi^\pm\jpsi)$. The
efficiency curve is considered in the fit and the possible
interference between the signal and background is neglected.
Figure~\ref{1Dfit} shows the fit results; the fit yields a mass of
$(3899.0\pm 3.6)~{\rm MeV}/c^2$, and a width of $(46\pm 10)$~MeV.
The goodness-of-the-fit is found to be $\chi^2/ndf=32.6/37=0.9$.

The number of $\X$ events is determined to be $N(\X^\pm)=307\pm
48$. The production ratio is calculated to be $R =
\frac{\sigma(\EE\to \pi^\pm \X^\mp\to \pcpcjpsi)}
{\sigma(\EE\to\pcpcjpsi)} = (21.5\pm 3.3)\%$, where the efficiency
correction has been applied. The statistical significance is
calculated by comparing the fit likelihoods with and without the
signal. Besides the nominal fit, the fit is also performed by
changing the fit range, the signal shape, or the background shape.
In all cases, the significance is found to be greater than
$8\sigma$.

Fitting the $M(\pip\jpsi)$ and $M(\pim\jpsi)$ distributions
separately, one obtains masses, widths, and production rates of
the $\X^+$ and $\X^-$ that agree with each other within
statistical errors. Dividing the sample into two different
$M(\pp)$ regions (below and above $M^2(\pp)=0.7~{\rm GeV}^2/c^4$)
allows us to check the robustness of the $\X$ signal in the
presence of two different sets of interfering $\pcpcjpsi$
amplitudes. In both samples, the $\X$ is significant and the
observed mass can shift by as much as $14\pm 5~{\rm MeV}/c^2$ from
the nominal fit, and the width can shift by $20\pm 11~{\rm MeV}$.
We attribute the systematic shifts in mass and width to
interference between the $\X\pi$ and $(\pp)\jpsi$ amplitudes. In
fitting the $\pi^\pm\jpsi$ projection of the Dalitz plot, our
analysis averages over the entire $\pp$ spectrum, and our
measurement of the $\X$ mass, width, and production fraction
neglects interference with other $\pcpcjpsi$ amplitudes.

The systematic errors for the resonance parameters of the $\X$
come from the mass calibration, parametrization of the signal and
background shapes, and the mass resolution. The uncertainty from
the mass calibration can be estimated using the difference between
the measured and known $\jpsi$ masses (reconstructed from $\EE$
and $\MM$) and $D^0$ masses (reconstructed from $K^-\pip$). The
differences are $(1.4\pm 0.2)$~MeV/$c^2$ and $-(0.7\pm
0.2)$~MeV/$c^2$, respectively. Since our signal topology has one
low momentum pion, as in $D^0$ decay, and a pair of high momentum
tracks from the $\jpsi$ decay, we assume these differences added
in quadrature is the systematic error of the $\X$ mass measurement
due to tracking. Doing a fit by assuming a P-wave between the $\X$
and the $\pi$, and between the $\jpsi$ and $\pi$ in the $\X$
system, yields a mass difference of 2.1~MeV/$c^2$, a width
difference of 3.7~MeV, and production ratio difference of 2.6\%
absolute. Assuming the $\X$ couples strongly with $D\bar{D^*}$
results in an energy dependence of the total width~\cite{flatte},
and the fit yields a difference of 2.1~MeV/$c^2$ for mass,
15.4~MeV for width, and no change for the production ratio. We
estimate the uncertainty due to the background shape by changing
to a third-order polynomial or a phase space shape, varying the
fit range, and varying the requirements on the $\chi^2$ of the
kinematic fit. We find differences of 3.5~MeV/$c^2$ for mass,
12.1~MeV for width, and 7.1\% absolute for the production ratio.
Uncertainties due to the mass resolution are estimated by
increasing the resolution determined by MC simulations by 16\%,
which is the difference between the MC and measured mass
resolutions of the $\jpsi$ and $D^0$ signals. We find the
difference is 1.0~MeV in the width, and 0.2\% absolute in the
production ratio, which are taken as the systematic errors.
Assuming all the sources of systematic uncertainty are
independent, the total systematic error is 4.9~MeV/$c^2$ for mass,
20~MeV for width and 7.5\% for the production ratio.

In Summary, we have studied $\EE\to \pcpcjpsi$ at a CM energy of
4.26~GeV. The cross section is measured to be $(62.9\pm 1.9 \pm
3.7)$~pb, which agrees with the existing results from the
BaBar~\cite{babarnew}, Belle~\cite{belley4260}, and
CLEO~\cite{cleoy4260} experiments. In addition, a structure with a
mass of $(3899.0\pm 3.6\pm 4.9)~{\rm MeV}/c^2$ and a width of
$(46\pm 10\pm 20)$~MeV is observed in the $\pi^\pm \jpsi$ mass
spectrum. This structure couples to charmonium and has an electric
charge, which is suggestive of a state containing more quarks than
just a charm and anti-charm quark. Similar studies were performed
in $B$ decays, with unconfirmed structures reported in the
$\pi^\pm\psip$ and $\pi^\pm\chi_{c1}$
systems~\cite{belle_z4430,babar_z4430,belle_z12,babar_z12}. It is
also noted that model-dependent calculations exist that attempt to
explain the charged bottomoniumlike structures which may also apply
to the charmoniumlike structures, and there were model predictions of
charmoniumlike structures near the $D\bar{D^*}$ and $D^*\bar{D^*}$
thresholds~\cite{models}.


The BESIII collaboration thanks the staff of BEPCII and the
computing center for their hard efforts. This work is supported in
part by the Ministry of Science and Technology of China under
Contract No. 2009CB825200; National Natural Science Foundation of
China (NSFC) under Contracts Nos. 10625524, 10821063, 10825524,
10835001, 10935007, 11125525, 11235011; Joint Funds of the
National Natural Science Foundation of China under Contracts Nos.
11079008, 11179007; the Chinese Academy of Sciences (CAS)
Large-Scale Scientific Facility Program; CAS under Contracts Nos.
KJCX2-YW-N29, KJCX2-YW-N45; 100 Talents Program of CAS; German
Research Foundation DFG under Contract No. Collaborative Research
Center CRC-1044; Istituto Nazionale di Fisica Nucleare, Italy;
Ministry of Development of Turkey under Contract No.
DPT2006K-120470; U. S. Department of Energy under Contracts Nos.
DE-FG02-04ER41291, DE-FG02-05ER41374, DE-FG02-94ER40823; U.S.
National Science Foundation; University of Groningen (RuG) and the
Helmholtzzentrum fuer Schwerionenforschung GmbH (GSI), Darmstadt;
National Research Foundation of Korea Grant No. 2011-0029457 and
WCU Grant No. R32-10155.


\end{document}